\documentclass[prd,10pt,nofootinbib,preprint,superscriptaddress,a4paper]{revtex4-1}

\pdfoutput=1 
\usepackage{mciteplus}
\usepackage{dirtytalk}

\usepackage{tikz,xcolor,hyperref}

\usepackage{amsmath, amssymb, amsthm, graphicx, epsfig, fancyhdr,epsfig, slashed, mathrsfs}

\usepackage{tikzsymbols}
\usepackage{natbib}
\usepackage{float}

\usepackage{mathtools} 
\newcommand{\ba}{\begin{eqnarray}}
\newcommand{\ea}{\end{eqnarray}}
\newcommand{\be}{\begin{equation}}
\newcommand{\ee}{\end{equation}}
\newcommand{\bi}{\begin{itemize}}
\newcommand{\ei}{\end{itemize}}

\begin{document}

\title{Energy conditions for regular black holes in EFT of gravity}

\author{Ziyue Zhu}
\email{zhuziyue25@mails.ucas.ac.cn}
\affiliation{School of Fundamental Physics and Mathematical Sciences, Hangzhou Institute for Advanced Study, UCAS, Hangzhou 310024, China}

\author{Alexey S. Koshelev}
\email{askoshelev@shanghaitech.edu.cn}
\affiliation{
School of Physical Science and Technology, ShanghaiTech University, 201210 Shanghai, China
}
\affiliation{
	Departamento de F\'isica, Centro de Matem\'atica e Aplica\c{c}oes (CMA-UBI),
	Universidade da Beira Interior, 6200 Covilh\~a, Portugal }

\author{Yang Liu}
\email{liuyang2023@shanghaitech.edu.cn}
\affiliation{
School of Physical Science and Technology, ShanghaiTech University, 201210 Shanghai, China
}

\author{Anna Tokareva}
\email{tokareva@ucas.ac.cn}
\affiliation{School of Fundamental Physics and Mathematical Sciences, Hangzhou Institute for Advanced Study, UCAS, Hangzhou 310024, China}
\affiliation{International Centre for Theoretical Physics Asia-Pacific, Beijing/Hangzhou, China}
\affiliation{Theoretical Physics, Blackett Laboratory, Imperial College London, SW7 2AZ London, U.K.}

\begin{abstract}
As Einstein's gravity is a non-renormalizable theory, it can be a good description of physics only at the scales of energy or spacetime curvature below the Planck mass. Moreover, it requires the presence of an infinite tower of higher-derivative corrections, as required in the framework of effective field theory (EFT). Black holes, known to be vacuum solutions in Einstein's gravity, necessarily have singularities in the center, where both Einstein's gravity and low-energy EFT expansions break down. In this work, we address the question of whether, in the presence of matter, regular solutions looking like black holes from outside do exist. We show that the matter distribution supporting the regular black hole solution in the presence of Riemann tensor cube and Riemann tensor to the fourth power EFT corrections satisfies positivity of energy (also called weak energy condition, WEC) and null energy condition (NEC) everywhere outside the horizon. Unlike the case of singular solutions, the EFT description is also valid in the interior of such an object, given that the maximal curvature is bounded and does not exceed the cut-off scale. We found that in a wide range of parameters, WEC is satisfied inside the horizon, but NEC is violated inside the horizon in all cases.
\end{abstract}

\maketitle
\clearpage

\section{Introduction}

Black holes (BH) are perhaps the most puzzling objects found upon studying Einstein's General Relativity (GR) \cite{Wald:1984rg}.
Schwarzschild metric \cite{Schwarzschild:1916uq} is the very first celebrated solution to GR equations of motion describing a BH --- a static compact centrally symmetric massive object with an event horizon in an empty space.
This metric has neatly described the Mercury perihelion displacement but has also brought to life the problem of a BH singularity. By singularity, we mean a point or a region in the space-time where a metric or a curvature is either singular or non-differentiable a required number of times \cite{Kerr:2023rpn}. Indeed, the Riemann tensor and the corresponding scalar invariants evaluated on the Schwarzschild metric are singular at the center of a BH. The appearance of singularities can naturally be related to the fact that GR, while being perfect for describing IR gravity effects, is not UV complete and thus fails at high energies. It is therefore a wishful expectation that a complete gravity theory avoids singularities, including the BH ones \cite{Koshelev:2024wfk}.

Constructing a non-perturbative gravity theory appears to be an extremely difficult problem that has captured huge efforts of many top-notch scientists for a century without a satisfactory outcome yet. However, this problem can be attacked via a bottom-up approach in the framework of an effective field theory (EFT) of gravity. EFT of gravity can be constructed by adding to the original Lagrangian of GR new terms containing higher curvature corrections. They appear as counterterms required to cancel loop divergences in GR, however, as GR is non-renormalizable, one needs an infinite number of them. A Riemann curvature factor is equivalent to a two-derivative operator, and thus higher curvature terms can probe higher energy regimes. One can straightforwardly add terms cubic and quartic in the Riemann tensor \cite{Ruhdorfer:2019qmk}.\footnote{Note that quadratic in curvature terms can be removed by a field redefinition.} An interesting question about how vacuum BH solutions of GR get modified within the EFT approach has been widely studied \cite{Boyce:2025fpr,Melville:2024zjq,Cano:2019ycn,Saraswat:2016eaz,Chen:2024sgx,Barbosa:2025uau,Kats:2006xp}. This, unfortunately, does not give an immediate insight into how a singularity problem can be resolved. The EFT corrections to the BH solutions are typically also singular; however, EFT breaks down near the singularity, so a complete theory of gravity is required in order to answer the question about the presence of a singularity in the center of a black hole.

The other possible path is to start by guessing about possible regular BH solutions. Several interesting proposals of regular BH-s have been developed \cite{Bardeen:1968qqq,Frolov:1988vj,Dymnikova:1992ux,Ayon-Beato:1998hmi,Bonanno:2000ep,Bronnikov:2000vy,Hayward:2005gi,Saini:2014qpa, Alencar:2023wyf,Ansoldi:2008jw,Aguayo:2025xfi} (see \cite{Lan:2023cvz} for a review). Such solutions were mainly constructed in a framework of GR. They are not vacuum solutions, and an extra matter content needs to be added. In general, regular BH-s prompt a very important question of how their interior is organized. The Hawking-Penrose theorems \cite{Penrose:1964wq,Hawking:1971vc} suggest that the presence of an event horizon leads to a singularity. To circumvent this statement, one has to modify the BH interior, for example, by allowing for an inner horizon. This, in turn, may lead to other problems, including the so-called mass inflation \cite{Visser:2024zkx,Gao:2025plm,Brown:2011tv} which is, in general, not easy to avoid. Moreover, adding extra matter requires checking that various reasonable energy conditions \cite{Zaslavskii:2010qz,Neves:2014aba,Lan:2022bld} are satisfied.

In the present paper, we combine the idea of the EFT of gravity and the construction of regular BH-s. Namely, we are going to embed regular BH candidate metrics in the EFT of gravity models. In particular, we consider cubic and quartic in Riemann tensor terms. We do not include parity-violating terms. Since the candidate metrics are not exact vacuum solutions, extra matter is required to validate these solutions. The idea of this work is to establish conditions on EFT coefficients such that matter-energy conditions are satisfied. These conditions may depend on how the interior of a BH is regularized, and thus different regular solutions have to be considered. Having a vacuum solution without extra matter is a nice goal that is beyond the scope of the present paper. The complexity of the field equations does not allow for constructing such solutions beyond the perturbative approach, which is commonly used in the EFT of gravity and is known to provide only singular solutions. The latter are not complete, as they predict the breakdown of EFT near the center of the black hole. The main goal of this paper is to construct a regular solution with finite maximal curvature that can be fully described within the regime of validity of EFT. We check whether this is possible with matter satisfying at least the weak energy condition both inside and outside the outer horizon of a refular BH.

The paper is organized as follows. In Section~\ref{sec:setup} we introduce the model to be examined in the rest of the paper. In Section~\ref{sec:constraintsEFT} we study constraints on suggested regular BH solutions coming from EFT considerations. In Section~\ref{constraintsMatter}
we study constraints originating from the viability of the matter content needed to support BH configurations in question. In Section~\ref{sec:conclusions} we put concluding remarks and formulate problems for future study directions. Appendix contains explicit expressions used for producing plots presented in the main text.

\section{The setup}
\label{sec:setup}

In what follows, we work in the EFT approach to gravity in four dimensions by accounting lowest order corrections to GR. That is, we consider cubic and quartic in curvature terms in the action as follows
\begin{equation}
    S = \int d^4 x\sqrt { - g}  \frac{M_p^2}{2}\left(R + \frac{C_1}{M_p^4}R_{abcd}R_{\phantom{ab}ef}^{ab}R^{cdef}+ \frac{C_2}{M_p^6}\left(R_{abcd}R^{abcd}\right)^2\right)\,,
    \label{actionmain}
\end{equation}
where $C_{1,2}$ are dimensionless constants. Here we omit and will not consider parity violating terms which involve a $D$-dimensional fully antisymmetric tensor density $\epsilon$.

Our goal is to study regular BH-s within the EFT approach and find out possible constraints on this setup. We use the term Black Hole to indicate an object that has an event horizon. In order to represent a regular space-time geometry, it should have at least two horizons, outer and inner. In principle, the considerations below can be equally applied to totally regular compact massive objects without any horizon.

There are two obvious sources for possible constraints. First, we want to ensure that the cut-off scales of the EFT are not exceeded by parameters and characteristic energy scales of the obtained solutions. Second, on the way of embedding black holes or, in general, compact massive objects in any gravity model, one may encounter a need to add some matter distribution in the exterior of this object to support this configuration. The simplest example is a charged Reissner-Nordstrom BH, which requires some configuration of the electromagnetic field around it to be a solution. It is crucial to check that the matter distribution satisfies viable energy conditions such as positivity of energy, or weak energy condition (WEC), and null energy condition (NEC), which will be explained below.


In our consideration of compact massive objects, we specialize to configurations without rotation. While in general rotation can be present, it will not change conceptually our study but will complicate the corresponding math considerably. For the moment, we defer a detailed consideration of EFT of gravity with rotating regular BH-s to future publications.

A spherically symmetric static metric in Schwarzschild coordinates can be written as follows
\begin{align}
    d{s^2} =  - f(r)d{t^2} + {\tilde f(r)^{ - 1}}d{r^2} + {r^2}\left( {{\rm{d}}{\theta ^2} + {{\sin }^2}\theta {\rm{d}}{\varphi ^2}} \right)\,,
    \label{SchMetric}
\end{align}
where the functions $f(r)$ and $\tilde f(r)$ can be different. Hereafter, we will mainly consider a simplified situation when $f(r)=\tilde f(r)$. The celebrated Schwarzschild metric is restored for
\begin{equation}
     f(r)=\tilde f(r) = 1 - \frac{{2GM}}{r}\,.
    \label{Schd}
\end{equation}
The Schwarzschild metric is a solution to the Einstein equations in an empty space with a point-like source of mass $M$. $G$ is the Newtonian constant related to the Planck mass as $8\pi G =1/M_p^2$. This solution nicely restores the Newtonian $\sim1/r$ potential far away from the gravitation source but leads to a singularity at the center. One can check that the Kretschmann invariant $K=R_{abcd}R^{abcd}$ is singular and behaves like $K\sim r^{-6}$ for small $r$.

A regular geometry can be yielded for a modified function $f(r)$. We should not change the long-range behavior to preserve an IR physics, but rather we can modify the short-range or UV behavior. We need at least to have $f(0)=1$ and $f'(0)=0$, where a prime is the derivative with respect to the radial coordinate $r$. We are saying ``at least'' because this is necessary to have all the terms in the Einstein equations of GR to be regular. That is, to make the Riemann tensor behave smoothly and thus generate a regular Kretschmann invariant, as well as the other higher curvature contractions of Riemann tensors. It may not be sufficient, though, for certain terms, especially those including higher derivatives of the curvature tensor in the action. To implement this idea, we introduce
\begin{align}
    f(r,\alpha ) = 1 - \frac{{2GM}}{r}A(r,\alpha )\,.
    \label{Modi_Schd}
\end{align}
Here $\alpha$ is a new parameter of dimension length, $A(r,\alpha)$ is a function such that $A(r,\alpha)/r\overset{r\to0}{\to} 0$, also $A(r,0)=1$, and moreover $A(r,\alpha)\overset{\alpha\to\infty}\to0$. Examples to be used in our analysis are
\begin{eqnarray}
A(r,\alpha ) &=& \frac{{{r^p}}}{{{r^p} + \alpha^p }}\text{ and }\label{regAfrac}\\
A(r,\alpha ) &=& {e^{ - {\alpha^p }/{r^p}}}\,,
\label{regAexp}
\end{eqnarray}
with $p$ some positive integer constant. While the first choice is more traditional and has been explored a lot \cite{Lan:2023cvz}, the second choice is motivated by applications to infinite derivative gravity theories, for which the derivatives up to all orders have to be regular everywhere \cite{Koshelev:2024wfk,Culetu:2015axa}.

There are two comments in order here. First, in gravity models with higher derivative terms in the action, we may need to have a more regular behavior of the metric at the origin. Namely, more derivatives of $f(r)$ are to be zero. This will guarantee that all terms in the equations of motion are regular, providing regular expressions for observables. This can be regulated by the parameter $p$ in the above examples for the function $A(r)$. Second, we should clearly state that we are talking about guesses for regular solutions and \textit{not} about a regularization. The latter is barely a way of doing computations and has to be removed at the end of the process, thereby raising a question of the regularization independence, while the former is a suggested space-time metric, and the new parameters appearing there become new physical scales.


Suggested modified metrics almost obviously do not satisfy the vacuum equations of motion of our model. Even though we would be thrilled to find a true vacuum solution, for the moment, this goal has not been achieved. It is known how to construct a vacuum solution as modifications to GR BH-s perturbatively in EFT \cite{Kats:2006xp,Barbosa:2025uau,Chen:2024sgx,Saraswat:2016eaz,Cano:2019ycn,Melville:2024zjq,Boyce:2025fpr} but those studies have not addressed the quest of finding regular geometries (if they are possible at all). A related issue is to understand whether a real singularity still persists because of an EFT breakdown above the cut-off scale.
We postpone finding a vacuum solution for regular compact massive objects to future publications and therefore proceed the other way. We can add matter to the system in order to support a regular geometry in modified Einstein equations. This matter content can be characterized by its stress-energy tensor $T_{ab}$, and the valid questions are whether the curvature of a solution does not exceed the EFT cut-off and simultaneously, the added matter satisfies various reasonable energy conditions. This will be examined in the next Section alongside other important questions.


\section{Constraints on curvature from EFT cut-off}
\label{sec:constraintsEFT}
We proceed by examining constraints on possible regular geometries coming from EFT considerations.

Recall that the first terms in the EFT expansion of the action for gravity are, 
\begin{equation}
S = \int {{d^4}} x\sqrt { - g} \left[ {\frac{{M_p^2}}{2}\left( {R + \frac{C_1}{{{M_p ^4}}}{R_{abcd}}{R^{ab}}_{ef}{R^{cdef}} + \frac{C_2}{{{M_p^6}}}{{\left( {{R_{abcd}}{R^{abcd}}} \right)}^2}} \right)} \right].
\end{equation}
This action can be the low-energy description for gravity, valid until a certain energy scale -- cut-off scale $\Lambda$. Einstein-Hilbert action is known to break down at the Planck scale, while higher-order terms can, in principle, cause breaking of the EFT expansion at a lower scale. A rough estimate of such a scale can be made from power-counting arguments, as a suppression scale for each operator. More accurately, this scale can be found from the unitarity breaking of the graviton scattering amplitude. However, as an order-of-magnitude estimate, power counting usually coincides with that. The scales $\Lambda_1,~\Lambda_2$ corresponding to Riem$^3$, Riem$^4$ terms, respectively, can be found as
\begin{equation}
    \frac{1}{{{\Lambda_1 ^4}}} = \frac{{{C_1}}}{{M_p^4}},\quad  \frac{1}{{{\Lambda_2 ^6}}} = \frac{{{C_2}}}{{M_p^6}}.
\end{equation}
Thus, the cut-off scale of the whole action is given by
\begin{equation}
\Lambda=\text{min}\left(M_p,~\Lambda_1,~\Lambda_2\right).
\end{equation}

For regular solutions, the curvature invariants of the spacetime are bounded. If the maximal values of curvatures do not exceed the EFT cut-off ($R<\Lambda^2$), the solution can be trusted everywhere, including the interior of a black hole. Indeed, for the regular metric \eqref{SchMetric} with
\begin{equation}
\label{PolyModi_Schd}
     f(r,\alpha ) = 1 - \frac{{2GM}}{r}A(r,\alpha )\,, \quad A(r,\alpha ) = \frac{{{r^p}}}{{{r^p} + \alpha^p }}
\end{equation}
we have Ricci scalar (here we specialize to $p=3$) 
 \begin{equation}
     R = \frac{{12M{\alpha ^3}\left( {2{\alpha ^3} - {r^3}} \right)}}{{{{M_p^2\left( {{\alpha ^3} + {r^3}} \right)}^3}}}.
 \end{equation}
The maximal value of this function is $24 M/(M_p^2\alpha^3)$, which is an estimate for the maximal curvature of the regular solution.
 The constraint $R_{max} \lesssim {\Lambda ^2}$, thus, can be written as 
\begin{equation}
\label{EFTcutoff}
 \frac{M}{{{\alpha ^3}}} < \frac{{{M_p}^4}}{{\sqrt {{C_1}} }}\ ,\quad\frac{M}{{{\alpha ^3}}} < \frac{{{M_p}^4}}{{\sqrt[3]{{{C_2}}}}}  \,.
\end{equation}

This condition is independent of $p$. For the other curvature invariants, the requirement ${R_{abcd}}{R^{abcd}} < {\Lambda ^4}$ yields the same condition as \eqref{EFTcutoff}. In the subsequent computations, we often set the units of ${M_p} = 1$, thus getting the conditions of validity of the EFT description in the form
\begin{equation}
    C_1 < {\alpha ^6}{M^{ - 2}}, \quad C_2<\alpha^9 M^{-3}.
\end{equation}

\section{Matter energy conditions}
\label{constraintsMatter}

In this Section, we explore constraints originating from the viability of the matter needed to support regular BH configurations.

Given that matter is present in the system, one may need to check whether energy conditions are satisfied \cite{Kontou:2020bta}. That is, the so-called Weak, Null, Strong, and Dominant energy conditions, which are mostly examined and are abbreviated as NEC, WEC, SEC, and DEC, respectively. In the present paper, we focus on WEC and NEC, as they are essential requirements for constructing a consistent description of a physical system. WEC has to be satisfied in any unitary theory, while NEC can still be violated in more complicated theories of matter, for example, gravity with multiple interacting fields. Denoting $T_{ab}$ the stress-energy tensor of matter, WEC can be formulated as the non-negativity condition $T_{ab}t^at^b\geq0$ for any future-oriented time-like vector $t^a$. NEC can be formulated as the non-negativity condition $T_{ab}k^ak^b\geq0$ for any null vector $k^a$.

For a diagonal metric, the stress-energy tensor is clearly diagonal with components $T^a_{b}={\rm diag}(-\rho, p_1, p_2, p_3)$. Otherwise one can use the orthonormal tetrad method \cite{Hawking:1973uf} to bring the stress-tensor to a form ${T_{ab }} = \rho e^0_a e^0_b  + {p_1}e^1_a e^1_b  + {p_2}e^2_a e^2_b  + {p_3}e^3_a e^3_b$. Here $e^\alpha_a$ is an orthonormal tetrad satisfying $g^{ab}e_a^\alpha e_b^\beta=\eta^{\alpha\beta}$ with $\eta^{\alpha\beta}$ to be the Minkowski metric.
Then we have the following energy conditions,
\begin{align}
    {\rm WEC}~:~& \rho  \ge 0,\rho  + {p_i} \ge 0,\text{ for }i = 1,2,3\label{WECeq}\\
    \label{NECeq}
    {\rm NEC}~:~&\rho  + {p_i} \ge 0,\text{ for }i = 1,2,3~.
\end{align}

\subsection{Weak Energy Condition}
We first examine the WEC formulated as $\rho  \ge 0$. Using as a somewhat simpler example of a fractional regularization (\ref{regAfrac}) and taking into account only the Riemann tensor cube term in the action, we get the following expression for energy in the case $p=3$
\begin{equation}
\begin{split}
  &\rho\left( {r,\alpha ,M,{C_1}} \right) =\frac{6 \alpha ^3 M}{{M_p}^2 (\alpha ^3+r^3)^2}+\frac{24C_1M^2}{{M_p}^{10} (\alpha ^3+r^3)^9}\times\\
  &\Big[2 M (-2 \alpha ^{18}+49 r^{18}-2034 \alpha ^3 r^{15}+18174 \alpha ^6 r^{12}
  -31120 \alpha ^9 r^9+12393 \alpha ^{12} r^6-870 \alpha ^{15} r^3)\\
  &-9 {M_p}^2 r (-80 \alpha ^{18}+5 r^{18}-211 \alpha ^3 r^{15}
  +1726 \alpha ^6 r^{12}-1362 \alpha ^9 r^9-2047 \alpha ^{12} r^6+1177 \alpha ^{15} r^3)\Big]
\end{split}
\label{rho for Riem3}
\end{equation}

\begin{figure}[h!]
        \centering
        \includegraphics[width=.6\textwidth]{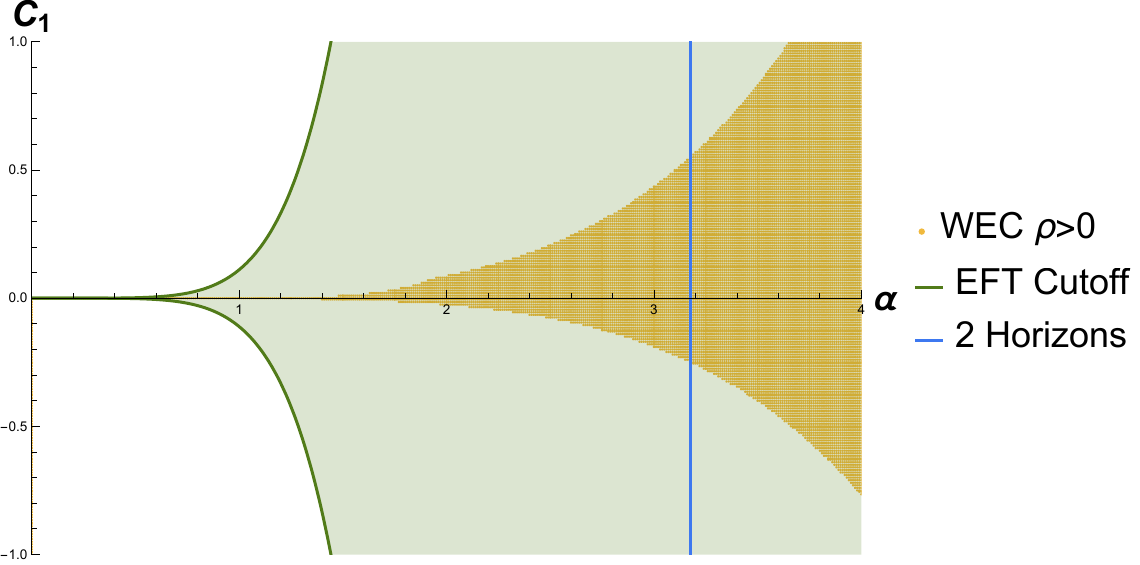 }
        \includegraphics[width=.6\textwidth]{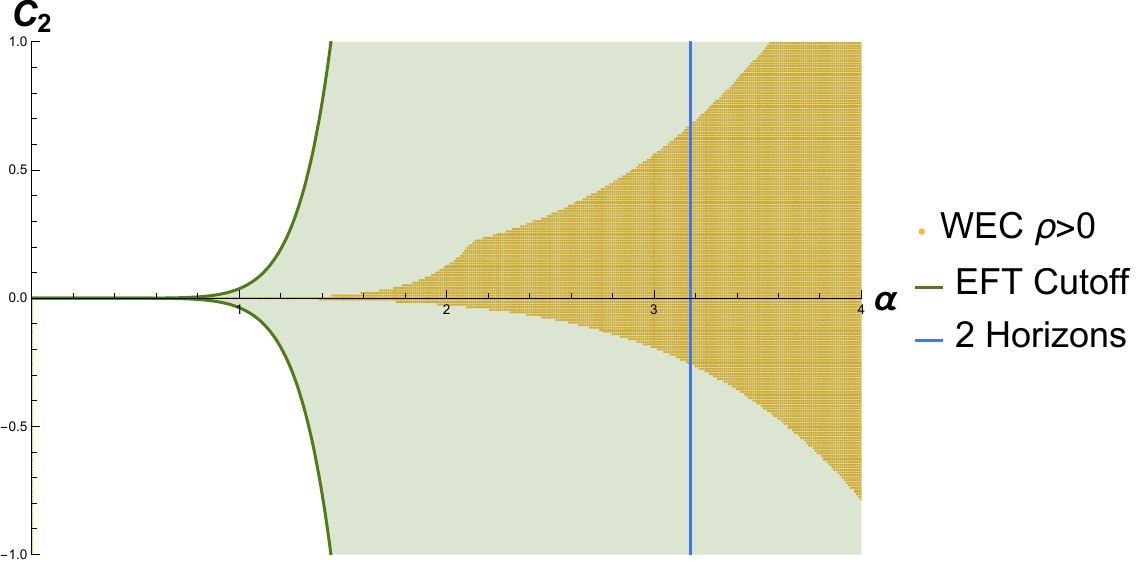 }
    \caption{This is a plot of the parameter space $C_{1,2}$ versus $\alpha$ computed for a fractional form of $A(r)$ (\ref{regAfrac}) in order to find parameter domains where an EFT does not break down and the WEC condition $\rho\ge0$ is satisfied. Here we fix $M=3,~p=3,~M_p=1$. The upper plot corresponds to $C_2=0$, i.e., only a Riemann tensor cube term is present. The lower plot corresponds to $C_1=0$, i.e, only the term quartic in the Riemann tensor is present. Within the green region, curvature does not exceed the cut-off scale of the EFT. Within the orange region, condition $\rho\ge0$ is satisfied. The blue line is the maximum value of $\alpha$ such that a horizon exists.}
    \label{WEC_sph}
\end{figure}

We can find regions in the parameter space $\{C_{1}, \alpha\}$ (for a fixed black hole mass $M$) such that for any value of the radius $r$, both the curvature limit discussed in the previous Subsection and the WEC for matter are satisfied. This is done numerically and depicted in the upper plot in Figure~\ref{WEC_sph}. The blue line indicates the maximum value of $\alpha$ for which two horizons are present. A similar analysis, but for a Riemann tensor to the fourth power, is done and depicted in the lower plot in Figure~\ref{WEC_sph}. The corresponding expression for the energy in the case $p=3$ is
\begin{equation}
\begin{split}
&\rho\left( {r,\alpha ,M,{C_2}} \right) =\frac{6 \alpha ^3 M}{{M_p}^2 (\alpha ^3+r^3)^2}+\frac{96 {C_2} M^3}{{M_p}^{14} (\alpha ^3+r^3)^{12}}\times\\& \Big[M (-8 \alpha ^{24}+303 r^{24}-14780 \alpha ^3 r^{21}+201850 \alpha ^6 r^{18}-974532 \alpha ^9 r^{15}+1379171 \alpha ^{12} r^{12}\\
&-617608 \alpha ^{15} r^9+88776 \alpha ^{18} r^6-3488 \alpha ^{21} r^3)\\
&-36 {M_p}^2 r (-40 \alpha ^{24}+4 r^{24}-194 \alpha ^3 r^{21}+2519 \alpha ^6 r^{18}-10423 \alpha ^9 r^{15}+5306 \alpha ^{12} r^{12}\\
&+10372 \alpha ^{15} r^9-6953 \alpha ^{18} r^6+1081 \alpha ^{21} r^3)\Big]
\end{split}
\end{equation}
In both plots, the yellow region corresponds to a situation when WEC is satisfied. It is a subdomain of a light green region, which shows when EFT remains valid. This designates a consistent subspace of parameters of the model and of a BH solution.
The same computation can be performed in the other coordinates for the purpose of verification of our results.
We repeated the above calculation with a Riemann tensor cube term in Eddington-Finkelstein coordinates, which yields a non-diagonal metric.\footnote{An explicit metric is $ds^2=-f\left( {r,\alpha } \right)d{u^2} + 2dudr + {r^2}\left( {{\rm{d}}{\theta ^2} + {{\sin }^2}\theta {\rm{d}}{\varphi ^2}} \right)$, where $du = dt + \frac{{dr}}{f\left( {r,\alpha } \right)}$.} 
{After rewriting the stress-energy tensor using tetrads, we find the energy density identical to \eqref{rho for Riem3} when $\alpha^3-2Mr^2/M_p^2+r^3>0$ is satisfied (for $p=3$), i.e., coordinate transformation works well in the exterior of the black hole where the coordinate transformation is well-defined.}

The same computations were performed in an exponentially regularized metric using a regulator factor (\ref{regAexp}). Despite having more bulky expressions for the energy, the final outcome is similar. Namely, there exists a non-empty parameter range such that both an EFT constraint on the maximal curvature and the WEC for matter are satisfied at the same time.

\subsection{Null Energy condition}
In order to study NEC, we have to consider whether an inequality $\rho  + {p_{{i}}} \ge 0$ is valid for all components of the pressure. For a static BH in Schwarzschild coordinates, we have to distinguish the pressure components: radial pressure $p_r=p_1$ and tangential pressure $p_a=p_2=p_3$.

\begin{figure}[h!]
        \centering
        \includegraphics[width=.6\textwidth]{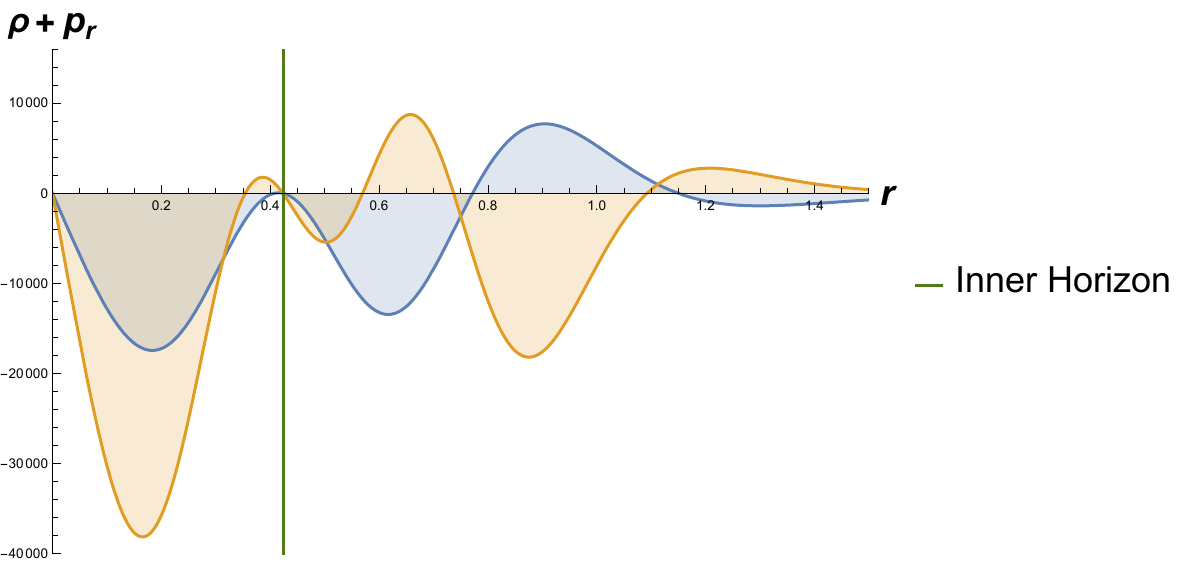}
        \centering
        \includegraphics[width=.6\textwidth]{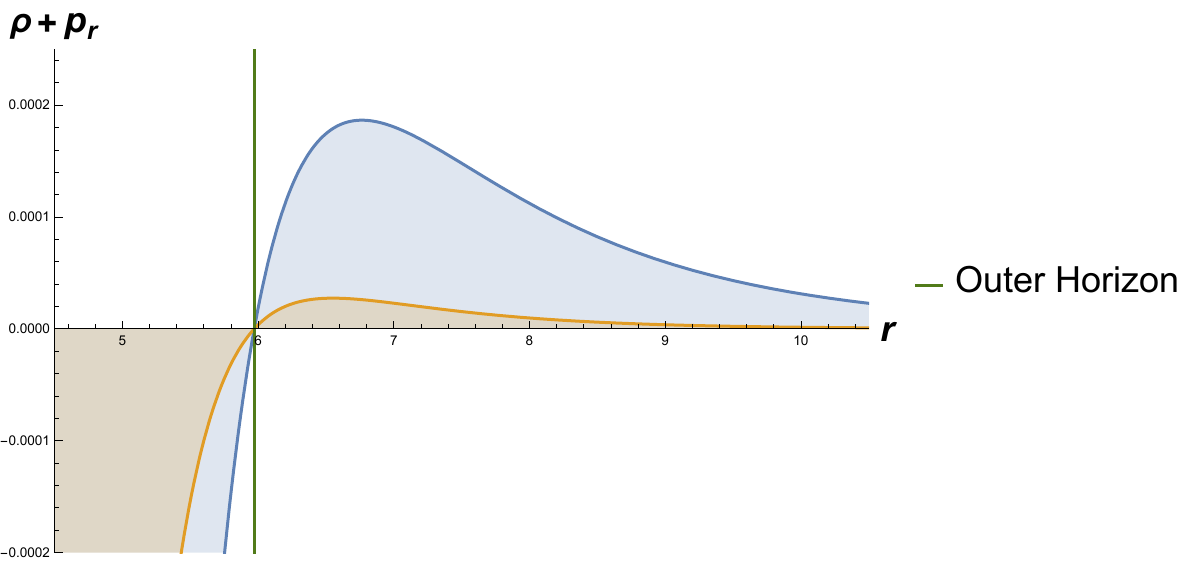}
        
      \caption{Here we plot quantity $\rho+p_r$ as a function of $r$, computed for a fractional form of $A(r)$ \eqref{regAfrac} where we fix $M=3,~p=3,~\alpha=3, M_p=1$. We split the whole range of $r$ for clarity. In the upper plot, the green line is the inner horizon and the EFT coefficients are taken $C_1=-1, C_2=0$ for the blue curve, and $C_1=0,  C_2= -0.1$ for the orange curve.
      In the lower plot, the green line is the outer horizon, and the EFT coefficients are taken as $C_1=-0.7,C_2=0$ for the blue curve, and $ C_1=0,C_2= -1$ for the orange curve.}      \label{NEC1}
    \end{figure}

    In GR, we always have $\rho  + {p_r}= 0$ such that the NEC is always satisfied irrespective of the function $f(r)$ given in (\ref{Modi_Schd}). This obviously is going to change upon adding new terms to the gravity action. Skipping explicit formulae which are cumbersome and not very much illuminating (see (\ref{rhopr}) for an analytic expression), we jump immediately to plots. In Figure~\ref{NEC1} we plot $\rho  + {p_r}$ as a function of the radius $r$ when the higher order terms in the Riemann tensor are added to the action. We find that $\rho+p_r $ always exhibits oscillating behavior under the outer horizon for both cubic and quartic in the Riemann tensor terms. In order to satisfy $\rho+p_r \ge 0$ outside the outer horizon, we have to set coefficients $C_{1,2}$ to be negative. Also, we see in Figure~\ref{NEC1} that $\rho+p_r$ becomes zero at the position of the horizon as it should be.

\begin{figure}[h!]
        \centering
        \includegraphics[width=.6\textwidth]{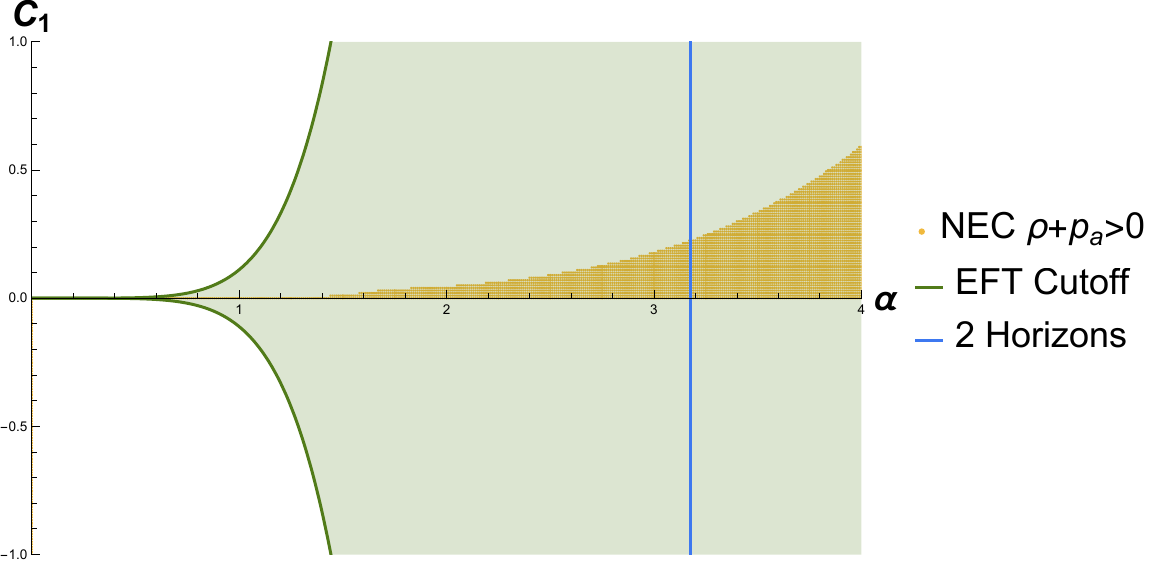}
        \includegraphics[width=.6\textwidth]{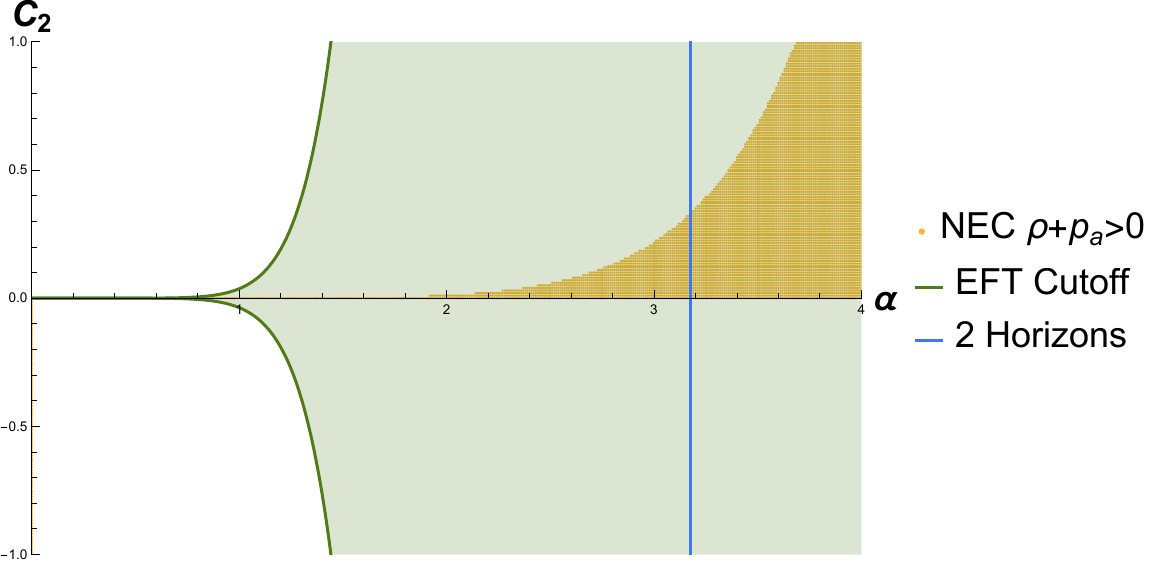}
     \caption{Here we plot the parameter space $C_{1,2}$ versus $\alpha$ computed for a fractional form of $A(r)$ (\ref{regAfrac})  in order to find parameter domains where an EFT does not break down and the NEC condition $\rho+p_a\ge0$ is satisfied everywhere. Here we fix $M=3,~p=3,~M_p=1$. The upper plot corresponds to $C_2=0$, i.e., only a Riemann tensor cube term is present. The lower plot corresponds to $C_1=0$, i.e, only the term quartic in the Riemann tensor is present. Within the green region, curvature does not exceed the cut-off scale of the EFT. Within the orange region, condition $\rho+p_a\ge0$ is satisfied. The blue line is the maximum value of $\alpha$ such that a horizon exists.
   }
     \label{NEC2}
    \end{figure}

\begin{figure}[h!]
        \centering
        \includegraphics[width=.6\textwidth]{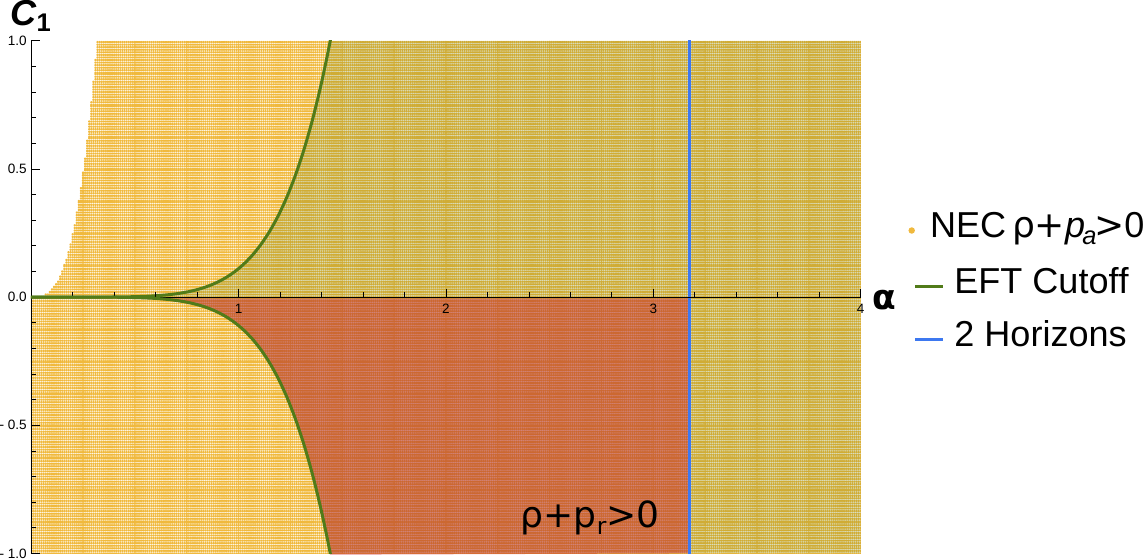}
        \centering
        \includegraphics[width=.6\textwidth]{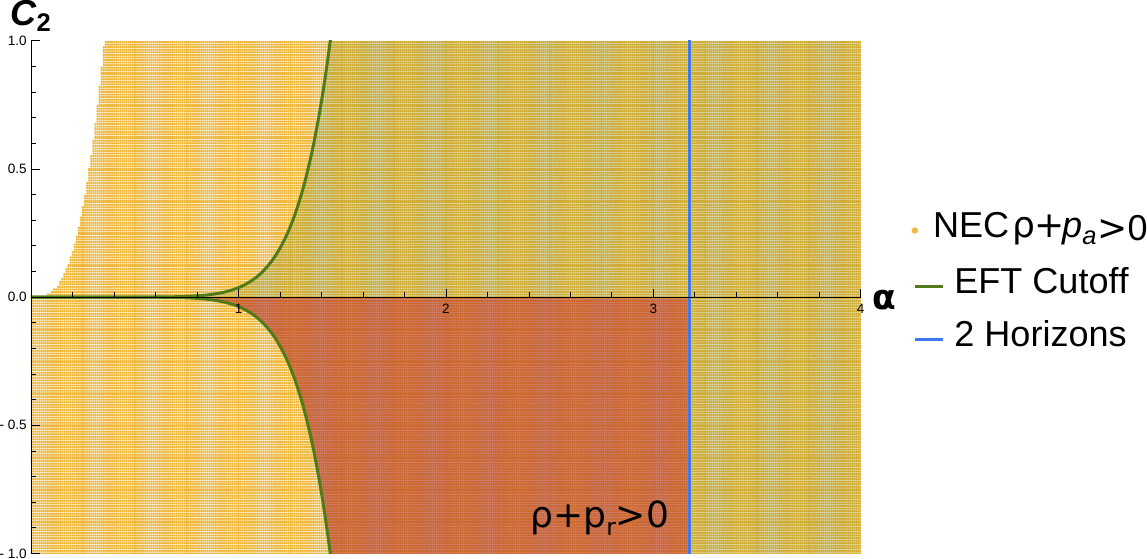}
        
      \caption{Here we plot the parameter space $C_{1,2}$ versus $\alpha$ computed for a fractional form of $A(r)$ (\ref{regAfrac}) in order to find parameters domains where an EFT does not break down and the NEC condition $\rho+p_a\ge0$ is satisfied in the exterior of a BH. Here we fix $M=3,~p=3,~M_p=1$. The upper plot corresponds to $C_2=0$, i.e., only a Riemann tensor cube term is present. The lower plot corresponds to $C_1=0$, i.e, only the term quartic in the Riemann tensor is present. Within the orange region, $\rho  + {p_a}\ge 0$ is satisfied outside the outer horizon. The red area in both plots shows the parameter domain, in which both radial and axial null energy conditions are satisfied outside the outer horizon.}      \label{NEC2 Outside}
    \end{figure}

    For the tangential pressure and the corresponding quantity $\rho+p_a$  (see (\ref{rhopa}) for an analytic expression), we have a sufficient positive contribution from the GR ter,m which can cancel oscillatory behavior originating from higher order Riemann tensor terms. This allows us to establish a region in the parameter space of $\{C_{1,2},\alpha\}$ like we did for the WEC. The corresponding plots are shown in Figure~\ref{NEC2}. It is easy to see that one has to have positive EFT coefficients $C_{1,2}$ in order to have condition $\rho+p_a\ge0$ for all $r$.
However, if one is concerned about the exterior of a BH only then the negative values of EFT parameters $C_{1,2}$ are also admissible. The corresponding situation is depicted in Figure~\ref{NEC2 Outside}.

Combining the results of the performed analysis of energy conditions, we see that negative EFT parameters are favored in the exterior of a static BH. However, one faces an imminent NEC violation both inside the inner horizon and between the horizons. While this situation may, in principle, change upon accounting for higher-order corrections to the gravitational action, it does not seem to be a big issue. NEC violation is not uncommon and can be easily found in other models. The most apparent example is an inflation with a single scalar field. Similar computations were performed for an exponentially regularized metric using (\ref{regAexp}) and the results are essentially very close, while more lengthy in computations.

\section{Conclusions}
\label{sec:conclusions}

Although the problem of the black hole singularity cannot be resolved without having an ultraviolet complete theory of gravity, the objects that resemble black holes with a horizon could be spherically symmetric solutions to equations of motion in a low-energy theory of gravity with matter --- an EFT of GR. Adjusting the matter distribution, one can straightforwardly get a regular solution with two horizons and no singularity in the center. As the curvature invariants are finite everywhere in the space for such solutions, they can be trusted in the regime of validity of the EFT. This is possible if the maximal values of the curvature invariants do not exceed the Planck scale. This is a constraint on the EFT parameters.

In this work, we addressed the question of whether the matter required for constructing such regular solutions satisfies several reasonable requirements, such as energy conditions. In this setup, these requirements can be checked both in the interior and exterior of the black hole, and the matter can be described in a way independent of the concrete UV completion for gravity. We found that in the presence of cubic and quartic terms in the Riemann tensor in the action, the matter can satisfy both WEC and NEC outside the outer horizon. The sufficient condition for this is the EFT coefficients $C_{1,2}$ are negative. We have also found that WEC can be respected everywhere for all values of $r$, including the interior of the regular black hole, while NEC is inevitably violated inside the outer horizon for any parameter choices studied in the paper. 
Thus, we obtained that the regular metric ansatzes can be promoted to a solution in the EFT of GR, with matter violating NEC only in the interior of a black hole. Additionally, such a solution doesn't break the EFT description of gravity.

There are several intriguing open problems that are left for future study. In particular, the presence of an inner horizon is known to cause, in some cases, instabilities, such as mass inflation.  Existing results were obtained mainly in GR, though they still indicate that there are ways to avoid this instability \cite{Gao:2025plm}. It has not yet been studied in detail whether this problem still remains in the EFT of GR. Maybe there are regular geometries that do not suffer from this problem in general. Next, the current study can be extended by including rotating compact massive objects, and also by considering metrics with different functions $f(r)$ and $\tilde f(r)$ in (\ref{SchMetric}). Another direction for future studies is the construction of regular vacuum solutions in non-perturbative gravity. However, it requires more assumptions about the complete theory of gravity, for example, asymptotic safety or matching to string theory in the UV.

\begin{acknowledgments}
The authors would like to thank Hong Lu and Giorgio di Russo for illuminating discussions.
The work of A.~T. was supported by the National Natural Science Foundation of China (NSFC) under Grant No. 1234710.
\end{acknowledgments}

\appendix
\section{Explicit expressions for NEC}

Expression for $\rho+p_r$ calculated for $p=3$ is as follows
\begin{align}
    \begin{split}
      &\rho+p_r=\frac{1296 {C_1} M^2 r}{{M_p}^{10} \left(\alpha ^3+r^3\right)^9}\left(2 M r^2-{M_p}^2 \left(\alpha ^3+r^3\right)\right)\times\\&\left(-12 \alpha ^{15}+r^{15}-36 \alpha ^3 r^{12}+324 \alpha ^6 r^9-548 \alpha ^9 r^6+213 \alpha ^{12} r^3\right) \\
         & +\frac{6912 {C_2} M^3 r}{{M_p}^8 \left(\alpha ^3+r^3\right)^{12}}(2 M r^2-{M_p}^2 \left(\alpha ^3+r^3\right))\times\\&(-18 \alpha ^{21}+2 r^{21}-98 \alpha ^3 r^{18}+1361 \alpha ^6 r^{15}-6568 \alpha ^9 r^{12}+9226 \alpha ^{12} r^9-4030 \alpha ^{15} r^6+567 \alpha ^{18} r^3)\,.
    \end{split}
    \label{rhopr}
\end{align}
Expression for $\rho+p_a$ calculated for $p=3$ is as follows
\begin{align}
\label{rhopa}
    \begin{split}
     & \rho+p_a= \frac{18 \alpha ^3 M r^3}{{M_p}^2 \left(\alpha ^3+r^3\right)^3}+\frac{432 {C_1} M^2 r}{{M_p}^{10} \left(\alpha ^3+r^3\right)^9}\times\Big[2 M r^2 (-37 \alpha ^{15}+r^{15}-105 \alpha ^3 r^{12}\\&+992 \alpha ^6 r^9-1757 \alpha ^9 r^6+672 \alpha ^{12} r^3)+{M_p}^2 (34 \alpha ^{18}-r^{18}+101 \alpha ^3 r^{15}-857 \alpha ^6 r^{12}\\&+735 \alpha ^9 r^9+1082 \alpha ^{12} r^6-578 \alpha ^{15} r^3)\Big]\\
       &+\frac{576 {C_2} M^3 r}{{M_p}^{14} \left(\alpha ^3+r^3\right)^{12}}\times\Big[M r^2 (-448 \alpha ^{21}+51 r^{21}-2608 \alpha ^3 r^{18}\\&+33546 \alpha ^6 r^{15}-161868 \alpha ^9 r^{12}+228667 \alpha ^{12} r^9-103596 \alpha ^{15} r^6+14508 \alpha ^{18} r^3)\\&-3 {M_p}^2 (-68 \alpha ^{24}+8 r^{24}-412 \alpha ^3 r^{21}+5005 \alpha ^6 r^{18}-20828 \alpha ^9 r^{15}+10546 \alpha ^{12} r^{12}\\&+20480 \alpha ^{15} r^9-14131 \alpha ^{18} r^6+2120 \alpha ^{21} r^3)\Big]\,.
    \end{split}
\end{align}


%

\end{document}